\documentclass[12pt,fleqn]{article}

\usepackage{amsfonts,amssymb,cite}
\usepackage{epsf,epsfig,graphics,graphicx}

\def\onecol{\onecolumn \mathindent 2em}

\def\noi{\noindent}

\newcommand{\Title}[1]{\noi {{\Large\bf #1}}\\[1ex]}

\def\Aunames#1{\noi{\bf #1}}
\def\auth#1{${}^{#1}$}
\def\Addresses#1{\medskip\noi \protect
    \begin{description}\itemsep -3pt {\it #1} \end{description}}
\def\addr#1#2{\item[${}^{#1}$]{\it #2}}

\newcommand{\Abstract}[1]{\vskip 2mm \begin{center}
        \parbox{16.4cm}{\small\noi #1} \end{center}\medskip}

\def\email#1#2{\footnotetext[#1]{e-mail: #2}\addtocounter{footnote}{1}}

\def\nqq{\hspace*{-2em}}
\def\nhq{\hspace*{-0.5em}}

\def\cm{\hspace*{1cm}}

\def\Jl#1#2{#1 {\bf #2},\ }

\def\ApJ#1 {\Jl{Astroph. J.}{#1}}
\def\CQG#1 {\Jl{Class. Quantum Grav.}{#1}}
\def\DAN#1 {\Jl{Dokl. AN SSSR}{#1}}
\def\GC#1 {\Jl{Grav. Cosmol.}{#1}}
\def\GRG#1 {\Jl{Gen. Rel. Grav.}{#1}}
\def\JETF#1 {\Jl{Zh. Eksp. Teor. Fiz.}{#1}}
\def\JETP#1 {\Jl{Sov. Phys. JETP}{#1}}
\def\JHEP#1 {\Jl{JHEP}{#1}}
\def\JMP#1 {\Jl{J. Math. Phys.}{#1}}
\def\NPB#1 {\Jl{Nucl. Phys. B}{#1}}
\def\NP#1 {\Jl{Nucl. Phys.}{#1}}
\def\PLA#1 {\Jl{Phys. Lett. A}{#1}}
\def\PLB#1 {\Jl{Phys. Lett. B}{#1}}
\def\PRD#1 {\Jl{Phys. Rev. D}{#1}}
\def\PRL#1 {\Jl{Phys. Rev. Lett.}{#1}}

\def\al{&\nhq}                         \def\d{\partial}
\def\lal{&&\nqq {}}                    \def\dst{\displaystyle}
\def\eq{Eq.\,}                         
\def\eqs{Eqs.\,}                       \def\fracd#1#2{{\dst\frac{#1}{#2}}}
\def\beq{\begin{equation}}             
\def\eeq{\end{equation}}               \def\Half{{\fracd{1}{2}}}
\def\bear{\begin{eqnarray}}            
\def\bearr{\begin{eqnarray} \lal}      
\def\ear{\end{eqnarray}}               
\def\earn{\nonumber \end{eqnarray}}    \def\kappa{\varkappa}

\def\nnn{\nonumber\\ \lal }            \def\diag{\mathop{\rm diag}\nolimits}
      
\def\yy{\\[5pt] {}}                    \def\const{{\rm const}}
                \def\eps{\varepsilon}
\def\eql{\al =\al}

\def\mn{_{\mu\nu}}
\def\MN{^{\mu\nu}}
\def\mN{_\mu^\nu}

\def\R{{\mathbb R}}

\def\wh{wormhole}
\def\whs{wormholes}
\def\bh{black hole}
\def\bhs{black holes}
\def\bu{black universe}
\def\bku{black-universe}
\def\bkus{black universes}
\def\sph{spherically symmetric}
\def\ssph{static, spherically symmetric}
\def\asflat{asymptotically flat}

\def\KS{Kantowski-Sachs}
\def\RN{Reissner-Nordstr\"om}
\def\dens{${\rm g/cm^3}$}

\begin{document}
\onecol

\Title{Magnetic black universes and wormholes\yy with a phantom scalar}

\Aunames {S.V. Bolokhov\auth{a,1}, K.A. Bronnikov,\auth{a,b,2}
            and M.V. Skvortsova\auth{a,b,3} }

\Addresses{
\addr a {Institute of Gravitation and Cosmology,
         PFUR, Miklukho-Maklaya St. 6, Moscow 117198, Russia  }
\addr b {Center of Gravitation and Fundamental Metrology,
         VNIIMS, Ozyornaya St. 46, Moscow 119361, Russia}
      }
\bigskip

\Abstract
 {We construct explicit examples of globally regular static, spherically
  symmetric solutions in general relativity with scalar and electromagnetic
  fields which describe traversable wormholes (with flat and AdS asymptotics)
  and regular black holes, in particular, black universes. A black universe
  is a nonsingular black hole where, beyond the horizon, instead of a
  singularity, there is an expanding, asymptotically isotropic universe. The
  scalar field in these solutions is phantom (i.e., its kinetic energy is
  negative), minimally coupled to gravity and has a nonzero self-interaction
  potential. The configurations obtained are quite diverse and contain
  different numbers of Killing horizons, from zero to four. This
  substantially widened the list of known structures of regular BH
  configurations. Such models can be of interest both as descriptions of
  local objects (\bhs\ and \whs) and as a basis for building nonsingular
  cosmological scenarios.
  }

\noi
UDK: 530.12, 524.882, 524.834 

\email 1 {bol-rgs@yandex.ru}
\email 2 {kb20@yandex.ru}
\email 3 {milenas577@mail.ru}

\section{Introduction}

  One of the basic problems of black hole (BH) physics is the existence of
  curvature singularities beyond the event horizons in the well-known
  Schwarzschild, Reissner-Nordstr\"om, Kerr and other solutions of general
  relativity and their analogs in other metric theories of gravity. For full
  understanding of BH physics and geometry it is highly desirable to get rid
  of singularities, and this is usually connected with the hopes for a
  future quantum gravity.  Still of great interest are attempts to construct
  non-singular BH models in the framework of classical gravity, and
  different classes of such models have been described in the literature.
  One of such classes, termed {\it black universes} \cite{pha1, pha4}, is in
  our view of particular interest since it combines the properties of
  wormholes (no center, and a regular minimum of the area of coordinate
  spheres), BHs (a Killing horizon separating static and non-static
  space-time regions) and non-singular cosmological models (at large times
  the non-static region reaches a de Sitter mode of isotropic expansion).
  The black universe models make possible a cosmological scenario where a
  phantom-dominated gravitational collapse in some ``mother'' universe
  creates our Universe whose expansion begins from a horizon, and the next
  stages are isotropization and de Sitter inflationary expansion. Other
  kinds of regular BHs discussed in the literature are classified in
  \cite{pha4}; see also the conclusion of the present paper.

  In the models described in \cite{pha1, pha4}, the material source of
  gravity is a phantom scalar field that differs from the canonical one by
  the sign of its kinetic energy. Such a field has been repeatedly discussed
  as a possible supporter of wormhole geometries (\cite{br73, h_ellis} and a
  great number of later papers; see \cite{wh-rev1, book-mifi} for recent
  reviews). The possible existence of phantom fields in the Nature is to a
  large extent favored by modern  cosmological observations, indicating that
  the accelerated expansion of our Universe may be caused by a dominating
  "dark energy" density with the pressure to density ratio $w$ smaller than
  -1.  One can note that values $w < -1$ seem to be not only admissible but
  even preferable for describing an increasing acceleration, as follows from
  the most recent estimates: $w = -1.10 \pm 0.14$ ($1\,\sigma$)
  \cite{komatsu} (according to the 7-year WMAP data) and $w =
  -1.069^{+0.091}_{-0.092}$ \cite{sullivan} (mainly from data on type Ia
  supernovae from the SNLS3 sample). In this connection, cosmological models
  with phantom scalar fields, i.e., those with a negative kinetic term, have
  gained considerable attention in the recent years (see, e.g.,
  \cite{cos-ph1, cos-ph2} and references therein). There are theoretical
  arguments both {\it pro et contra\/} phantom fields, and the latter seem
  somewhat stronger, see, for instance, a discussion in \cite{br-don10}.

  In this paper, accepting the existence of a phantom scalar as a working
  hypothesis, we would like to discuss new features of \wh\ and \bku\
  configurations which appear if, in addition to a scalar field, an
  electromagnetic field is invoked as a source of gravity.  As in
  \cite{pha1, pha4}, we deal with \ssph\ space-times, therefore the only
  kinds of electromagnetic fields are a radial electric (Coulomb) field and
  a radial magnetic (monopole) field. It should be stressed that in the
  latter case it is not necessary to assume the existence of magnetic
  charges (monopoles): in both \whs\ and \bkus\ a monopole magnetic field
  can exist without sources due to a specific space-time geometry. In the
  \wh\ case it perfectly conforms to Wheeler's idea of a ``charge without
  charge'' \cite{wheeler}: electric or magnetic lines of force simply thread
  the \wh.  In the case of a \bu, the picture is different on different
  sides: in the static region a possible observer sees a \bh\ with an
  electric or magnetic charge; in the cosmological region, this corresponds
  to a homogeneous primordial electric or magnetic field. For definiteness,
  we will speak of magnetic fields.

  One of motivations for the present study was that modern observations
  testify to a possible existence of a global magnetic field up to
  $10^{-15}$ Gauss, causing correlated orientations of sources remote from
  each other \cite{magn}, and some authors point out the possible primordial
  nature of such a magnetic field.

  The paper is organized as follows. In Section 2 we present the basic
  equations and make some general observations. In Section 3 we obtain
  explicit examples of \wh, \bku\ and other regular \bh\ solutions using the
  inverse problem method. Section 4 contains a discussion and, in
  particular, some numerical estimates concerning the possible magnetic
  field strength at different stages of the cosmological evolution.

\section{Basic equations}

  We consider the action\footnote
      {We choose the metric signature ($+,-,-,-$), the units $c = \hbar =
       8\pi G=1$, and the sign of $T\mN$ such that $T^0_0$ is the energy
       density.}
\beq                                                         \label{S}
      S = \Half \int \sqrt{-g} d^4 x \Big[
        R + 2 \eps g\MN \d_\mu\phi\d_\nu\phi - 2V(\phi) - F\mn F\MN \Big],
\eeq
  where $R$ is the scalar curvature, $g = \det (g\mn)$, and $F\mn$ is the
  electromagnetic field tensor, $\eps =+1$ corresponds to a normal scalar
  field $\phi$ and $\eps = -1$ to a phantom one.

  The general \ssph\ metric can be written in the form
\beq
     ds^2 = A(u) dt^2 - \frac{du^2}{A(u)} - r^2(u)d\Omega^2,  \label{ds2}
\eeq
  where we are using the so-called quasiglobal gauge $g_{00} g_{11} = -1$;
  $A(u)$ is called the redshift function and $r(u)$ the area function;
  $d\Omega^2 = (d\theta^2 + \sin^2\theta\, d\varphi^2)$ is the linear
  element on a unit sphere. The metric is only formally static: it is really
  static if $A > 0$, but it describes a \KS\ type cosmology if $A < 0$, and
  $u$ is then a temporal coordinate. In cases where $A$ changes its sign,
  regions where $A > 0$ and $A<0$ are called R- and T-regions, respectively.

  Let us specify which kinds of functions $r(u)$ and $A(u)$ are required for
  the metric (\ref{ds2}) to describe a \wh\ or a \bu.

\begin{enumerate} \itemsep 1pt
\item
     The range of $u$ should be $u \in \R$, where both $A(u)$ and $r(u)$
     should be regular, $r >0$ everywhere, and $r\to \infty$ at both ends.

\item
     A flat, de Sitter or AdS asymptotic behavior as $u\to \pm \infty$.

\item
     In the \wh\ case, absence of horizons (zeros of $A(u)$), and flat or AdS
     asymptotics at both ends.

\item
     In the \bku\ case, a flat or AdS asymptotic at one end and a de Sitter
     asymptotic at the other.
\end{enumerate}

  The existence of two asymptotic regions with $r \sim |u|$ (by item 2)
  requires at least one regular minimum of $r(u)$ at some $u=u_0$, at which
\beq
     r = r_0 >0, \cm r' =0, \cm r''> 0,                   \label{min}
\eeq
  where the prime stands for $d/du$. (In special cases where $r''=0$ at the
  minimum, we inevitably have $r''> 0$ in its neighborhood.)

  The necessity of violating the weak and null energy conditions at such
  minima follows from the Einstein equations. Indeed, one of them reads
\beq
      2A\, r''/r = -(T^t_t - T^u_u),                   \label{01comm}
\eeq
  where $T\mN$ are components of the total stress-energy tensor (SET).

  In an R-region ($A > 0$), the condition $r''>0$ implies $T^t_t -T^u_u <
  0$; in the usual notations $T^t_t = \rho$ (density) and $-T^u_u = p_r$
  (radial pressure) it is rewritten as $\rho + p_r < 0$, which manifests
  violation of the weak and null energy conditions. It is the simplest proof
  of this well-known violation near a throat of a \ssph\ \wh\
  (\cite{m-thorne}; see also \cite{book-mifi}).

  However, a minimum of $r(u)$ can occur in a T-region, and it is then not a
  throat but a bounce in the evolution of one of the \KS\ scale factors (the
  other scale factor is $[-A(u)]^{1/2}$). Since in a T-region $t$ is a
  spatial coordinate and $u$ temporal, the meaning of the SET components is
  $-T^t_t = p_t$ (pressure in the $t$ direction) and $T^u_u = \rho$;
  nevertheless, the condition $r'' > 0$ applied to (\ref{01comm}) again
  leads to $\rho + p_t < 0$, violating the energy conditions. In the
  intermediate case where a minimum of $r(u)$ coincides with a horizon
  ($A=0$), the condition $r'' > 0$ holds in its vicinity, along with all its
  consequences. Thus the energy conditions are violated near a minimum of
  $r$ in all cases.

  In what follows, we will assume that the space-time is \asflat\ as $u\to
  \infty$ and consider different behaviors of the metric as $u\to -\infty$.

  The scalar field $\phi(u)$ involved in the action (\ref{S}) in a
  space-time with the metric (\ref{ds2}) has the SET
\bearr
     T\mN [s] = \eps A(u) \phi'(u)^2 \diag (1,\ -1,\ 1,\ 1)
        + \delta\mN V(u).                                    \label{SET-s}
\ear
  The solutions of interest for us correspond to $\eps = -1$ but we preserve
  both values of $\eps$ in the equations for generality.

  The electromagnetic field compatible with the metric (\ref{ds2}) can have
  the following nonzero components:
\[
       F_{01} = - F_{10}\ {\rm (electric),\ and }\quad
       F_{23} = - F_{32}\ {\rm (magnetic),}
\]
  such that
\beq                                                          \label{F_mn}
       F_{01}F^{01} = -q_e^2/r^4(u),\cm
       F_{23}F^{23} = q_m^2/r^4(u),
\eeq
  where the constants $q_e$ and $q_m$ have the meaning of electric and
  magnetic charges, respectively. The corresponding SET is
\beq
     T\mN [e] = \frac{q^2}{r^4(u)} \diag (1,\ 1,\ -1,\ -1),
     \qquad    q^2 = q_e^2 + q_m^2.
\eeq

  Thus the electromagnetic field equations have already been solved in a
  general form, and we are left with the set of Einstein and scalar field
  equations. It can be written as follows:
\bear
      2(A r^2 \phi')'  \eql \eps r^2 dV/d\phi,          \label{e-phi}
\yy
              (A'r^2)' \eql - 2r^2 V + 2q^2/r^2,               \label{00}
\yy
                 r''/r \eql - \eps {\phi'}^2,                  \label{01}
\yy
         A (r^2)'' - r^2 A'' \eql 2 -4 q^2/r^2,                \label{02}
\yy                                                            \label{11}
      -1 + A' rr' + Ar'^2 \eql r^2 (\eps A \phi'^2 -V)- q^2/r^2,
\ear
  The scalar field equation (\ref{e-phi}) follows from
  (\ref{00})--(\ref{02}), which, given the potential $V(\phi)$, form a
  determined set of equations for the unknowns $r(u)$, $A(u)$, $\phi(u)$.
  \eq (\ref{11}) (the ${1\choose 1}$ component of the Einstein equations),
  free from second-order derivatives, is a first integral of
  (\ref{e-phi})--(\ref{02}) and can be obtained from (\ref{00})--(\ref{02})
  by excluding second-order derivatives. Moreover, \eq(\ref{02}) can be
  integrated giving
\bear
          r^4 B'(u) \equiv r^4 \biggl(\frac{A}{r^2}\biggr)' = - 2u
                             + 4q^2 \int \frac{du}{r^2(u)}.   \label{B'}
\ear
  where $B(u) \equiv A/r^2$.

  Let us note that \eqs (\ref{e-phi})--(\ref{11}) in the case of a massless
  scalar field $\phi$ have been solved long ago, in \cite{penney} for
  $\eps=+1$ and in \cite{br73} for $\eps = -1$. At $\eps = +1$ all such
  solutions possess a central singularity; with a phantom scalar
  ($\eps = -1$), there are both singular solutions and twice \asflat\ \whs\
  \cite{br73} but nothing like \bkus.

  We here seek solutions with a nonzero potential $V(\phi)$. It is known
  \cite{vac1} that \eqs (\ref{e-phi})--(\ref{11}) lead to a very narrow
  choice of possible global space-time structures in the case $q = 0$.
  Indeed, due to (\ref{02}), if $q=0$, the function $B(u)$ cannot have a
  regular minimum, therefore it can have at most two zeros (which coincide
  with zeros of $A(u)$ and hence correspond to horizons), and if the model
  is \asflat, say, at large $u$, only a single simple horizon is possible.
  We shall see how a nonzero charge $q$ changes the situation.

\section{Some particular models}

\subsection{Solutions}

  If one specifies the potential $V(\phi)$, it is, in general, very hard to
  solve the field equations. Alternatively, to find examples of solutions
  possessing particular properties, one may employ the inverse problem
  method, choosing some of the functions $r(u)$, $A(u)$ or $\phi(u)$ and
  then reconstructing the form of $V(\phi)$. We will do so, choosing a
  function $r(u)$ that can provide \wh\ and \bku\ solutions. Given $r(u)$
  and the charge $q$, the function $A(u)$ is found from (\ref{B'}) and
  $V(u)$ from (\ref{00}). Furthermore, $\phi(u)$ is found from (\ref{01}),
  and, as long as $r''/r \ne 0$, we obtain a monotonic function $\phi(u)$
  which then yields an unambiguous function $V(\phi)$.

  A simple example of the function $r(u)$ compatible with the requirements
  1--4 is \cite{pha1}
\beq                                            \label{r}
         r(u) = \sqrt{u^2 + b^2} = b \sqrt{x^2 +1}
\eeq
  where $x = u/b$, and $b > 0$ is an arbitrary constant (the length scale).
  Evidently, $r''(x) > 0$, as required, and thus we automatically fix
  $\eps=-1$; we also have $r \approx b|x|$ at large $|x|$.

  Let us formally put $b=1$, which will actually mean that the length scale
  is arbitrary but the quantities $r$, $q$, $m$ (the Schwarzschild mass in
  our geometrized units) etc., with the dimension of length, are expressed
  in units of $b$, the quantities $B$, $V$ and others with the dimension
  ${\rm (length)^{-2}}$ in units of $b^{-2}$, etc.; the quantities $A$ and
  $\phi$ are dimensionless.

  Now, the expression for $B' = B'(x)$ can be written as
\beq                                                         \label{B'x}
    B'(x) = \frac{2}{(1+x^2)^2} (p - x + 2q^2 \arctan x),
\eeq
  where $p$ is an integration constant. Further integration gives
\bearr                                                       \label{B}
       B(x) = B_0 + \frac{1+ q^2 + px}{1 + x^2}
       + \left(p + \frac{2q^2x}{1+x^2}\right)\arctan x + q^2\arctan^2 x,
\ear
  where $B_0$ is one more integration constant.

  Now suppose that our system is \asflat\ at $x\to +\infty$. Since $B=A/r^2$
  and $A\to 1$ at infinity, we require $B\to 0$ as $x\to \infty$ and thus
  fix $B_0$ as
\beq                                                        \label{B_0}
        B_0 = -\pi p/2 - \pi^2 q^2/4.
\eeq
  Furthermore, comparing the asymptotic expression $A = 1- 2m/x + o(x)$ for
  $A(x)$ with what is obtained from our expression for $A = B r^2$, we find
  a relation between the Schwarzschild mass $m$ and our parameters $p$ and
  $q$:
\beq                                                        \label{p-m}
        p = 3 m - \pi q^2.
\eeq
  Thus $B$ is a function of $x$ and two parameters, the mass $m$ and the
  charge $q$.

  Now we know the metric completely, while the remaining quantities $\phi(x)$
  and $V(\phi(x))$ are easily found from \eqs (\ref{01}) and (\ref{00}),
  respectively:
\bear                                             \label{phi_x}
    \phi(x) \eql \pm \arctan x + \phi_0,\cm \phi_0 = \const;
\yy                                                         \label{V_x}
    V(x) \eql \frac{q^2}{(1+x^2)^2}-\frac{1}{2(1+x^2)}
            \biggl[ 2q^2 (3x^2 + 1) \arctan^2 x
\nnn \ \ \
    + (18 x^2 m - 6\pi x^2q^2 + 12xq^2 - 2\pi q^2 + 6m) \arctan x
\nnn \ \ \
    + q^2\biggl(\frac 32\pi^2 x^2 - 6\pi x + \frac 12\pi^2 + 6\biggr)
        - m (9\pi x^2 - 18x +3\pi)\biggr].
\ear

  Thus $\phi$ has a finite range: $\phi \in (- \pi/2, \pi/2)$
  (putting $\phi_0 = 0$ without loss of generality), which is common to kink
  configurations.  We also have $x = u/b = \tan\phi$, whose substitution
  into the expression for $V(x)$ gives $V(\phi)$ defined in this finite range.

  It is easy to verify that asymptotic values of the function $B(x)$ at
  $x\to -\infty$ are directly related to those of the potential $V$ which in
  this case plays the role of an effective cosmological constant:
\beq
        V(-\infty) = -3 B(-\infty),                       \label{as-V}
\eeq
  so that negative $B(-\infty)$ correspond to a de Sitter (dS) asymptotic,
  with $B(-\infty)=0$ it is flat and with $B(-\infty)>0$ it is anti-de
  Sitter (AdS). The solutions obtained may be classified by this asymptotic
  behavior and by the number and nature of horizons appearing there. The
  latter correspond to regular zeros of the function $B(x)$. It turns out
  that inclusion of the electromagnetic field makes the solutions much more
  diverse than it was found previously for purely scalar-vacuum
  configurations \cite{pha1, pha4, trap1, trap2}.

\subsection {Symmetric configurations}

\begin{figure}[t]
\centering
    {\includegraphics[width=8cm]{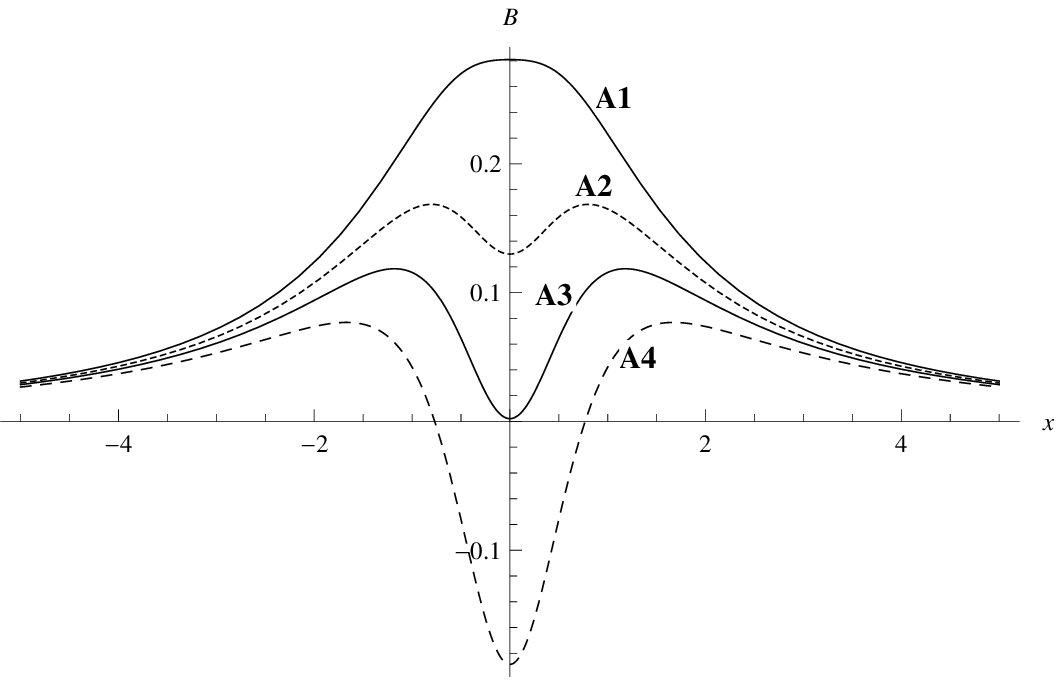}}\
    {\includegraphics[width=7cm]{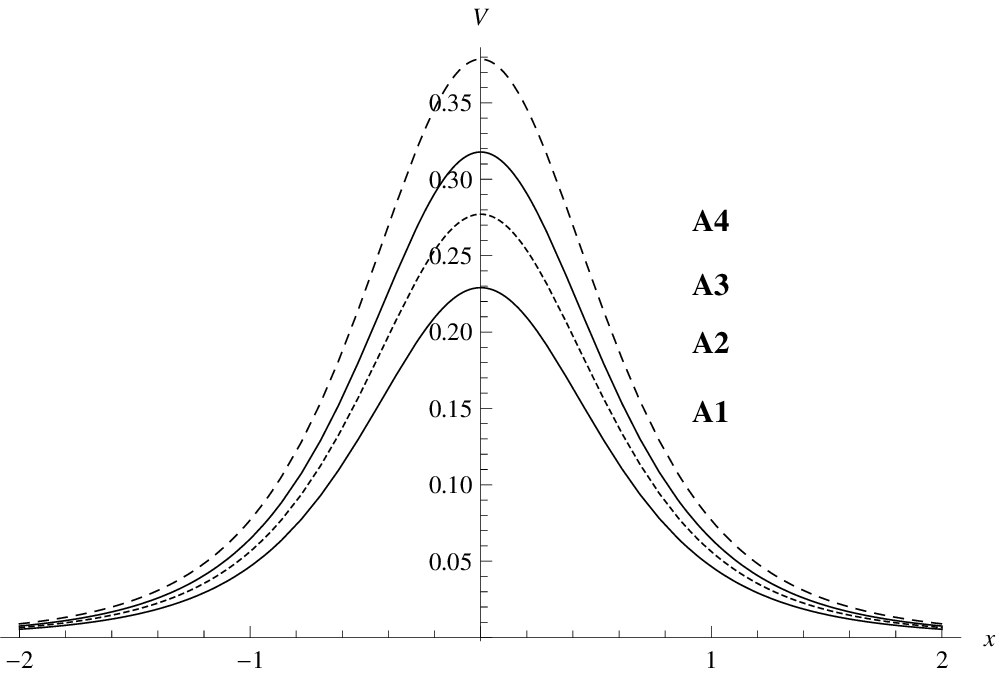}}\\[3mm]
    {\includegraphics[width=7cm]{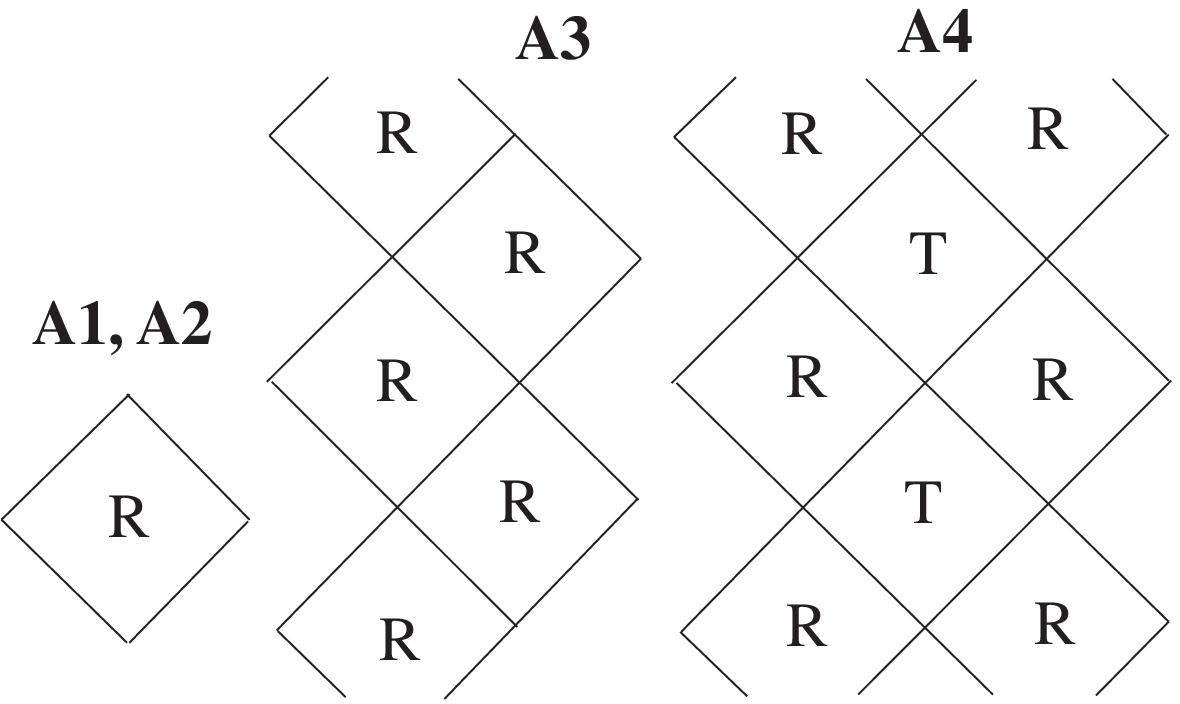}}
\caption{\small Plots of $B(x)$ (left) and $V(x)$ (right) for
    symmetric configurations. Curves A1, A2, A3, A4 correspond to
    $q = 0.7,\ 0.77,\  0.82461,\ 0.9$,
    respectively. The lower panel shows the Carter-Penrose diagrams of the
    corresponding space-times.  }
\label{fig-BVsym}
\end{figure}
  To begin with, from (\ref{B}) it follows that $B(x)$ is an even function
  if and only if $p=0$, hence $m = (2/3)\pi q^2$. Then $V(x)$ is also an
  even function. Such symmetric configurations are \asflat\ at both ends,
  $x \to \pm \infty$, and can be classified as follows (see the corresponding
  curves in Fig.\,1):
\begin{description}
\item [A1, A2:]
    Twice \asflat\ (M-M) \whs. The curve A2 contains a minimum of $B(x)$ at
    $x = 0$.
\item [A3:]
    Extremal regular \bhs\ (M-M), with a double horizon (curve A2).
\item [A4:]
    Non-extremal regular \bhs\ (M-M), with two simple horizons (curve A3).
\end{description}
  The abbreviation (M-M) stands here for two flat (Minkowski) asymptotic
  regions; we will also use similar notations for de Sitter (dS) and anti-de
  Sitter (AdS) asymptotic behaviors.

  The symmetric models form a one-parameter family, depending on $q$;
  clearly, at $q$ smaller than those appearing in Fig.\,1 we also obtain
  \whs\ (the simplest of them is with $q = m = 0$ and $V\equiv 0$, it is the
  Ellis massless \wh\ \cite{h_ellis, br73}), while at larger $q$ there are
  non-extremal regular \bhs. The critical value of $q$ that separates them is
  $q \approx 0.825$ (and $m \approx 0.713$), at which there emerges a
  double horizon corresponding to a double root of $B(x)$, hence a regular
  extremal BH.

\begin{figure}
\centering
      {\includegraphics[width=8cm]{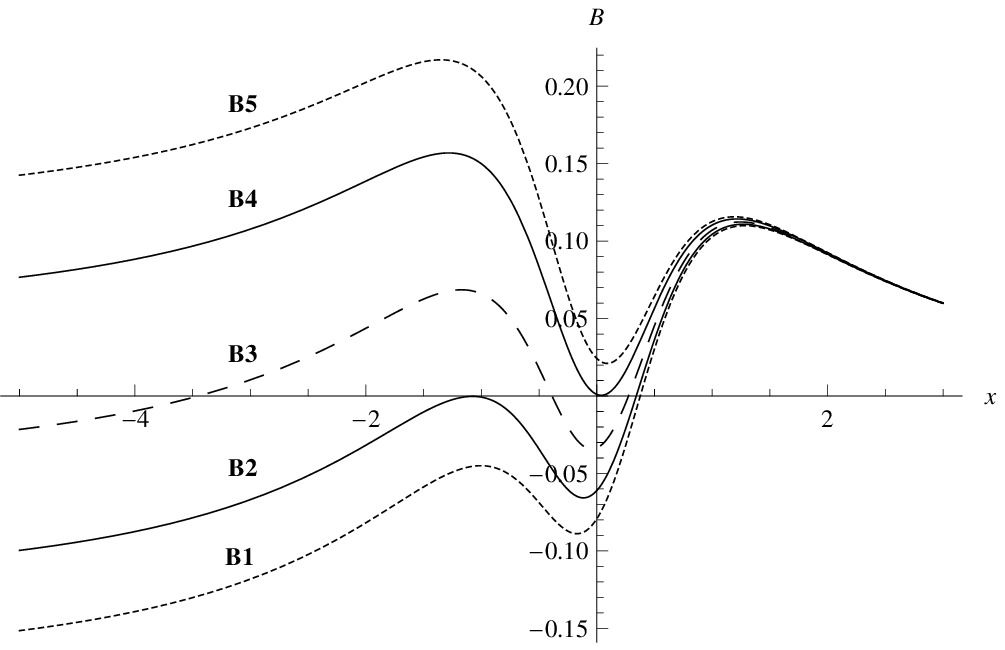}}\ \ \
      {\includegraphics[width=8cm]{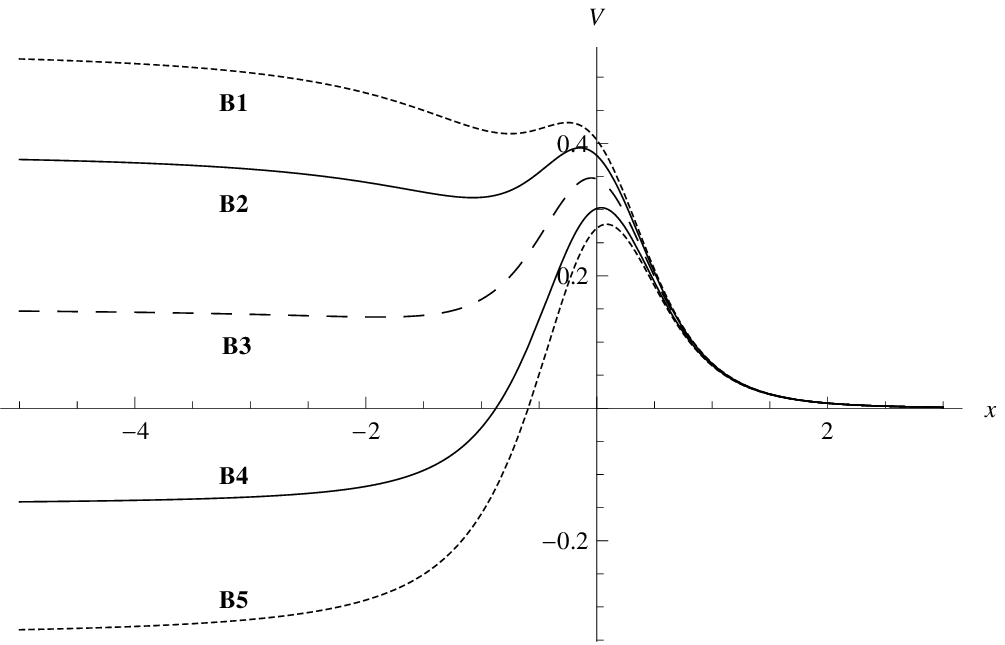}} \\[8pt]
      \raisebox{7mm}{\includegraphics[width=5cm]{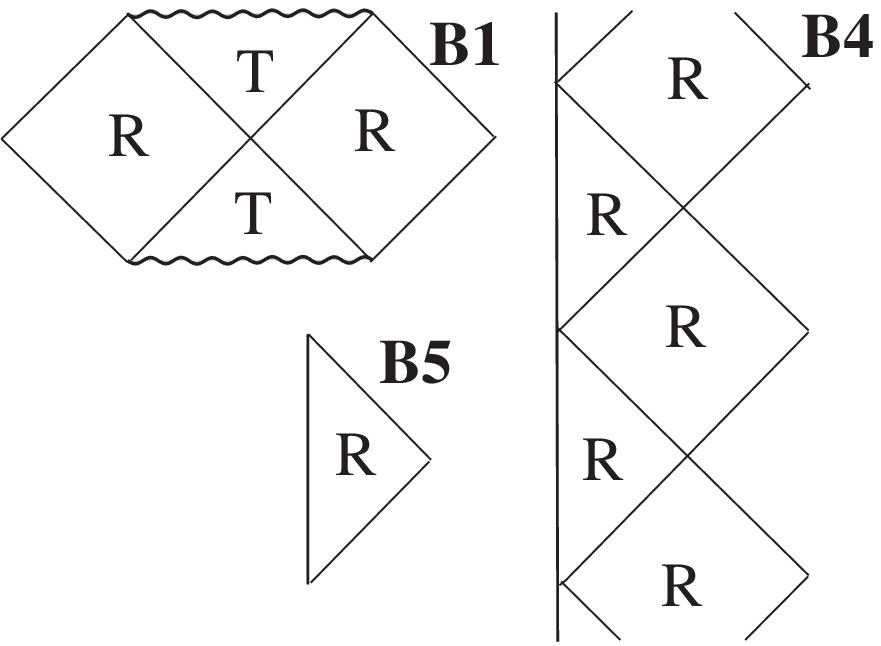}}\qquad
      {\includegraphics[width=5cm]{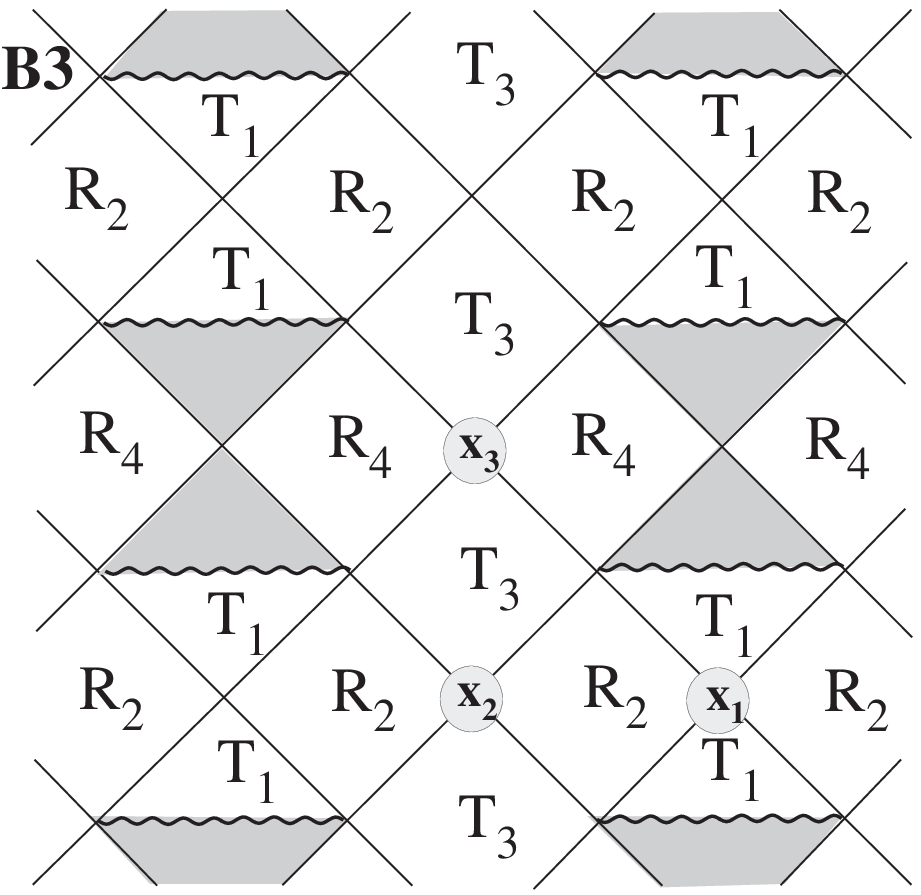}}\\[8pt]
      {\includegraphics[width=8cm]{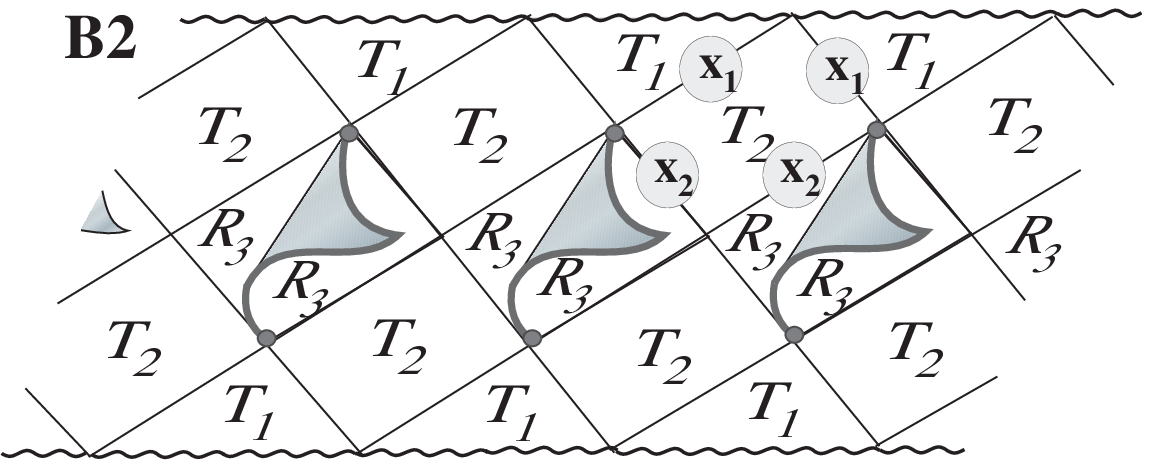}}
\caption{\small Plots of $B(x)$ (top, left) and $V(x)$ (top, right) for
    asymmetric configurations at parameter values close to critical ones.
    The parameters are:  $m= 0.725$ and $q = 0.821,\ 0.8242,\ 0.829,\
    0.835,\ 0.839$ for curves B1--B5, respectively.
    In the corresponding Carter-Penrose diagrams, the letters R and T label
    R- and T-regions, respectively, while indices near R and T enumerate
    ranges with a certain sign of $B(x)$ along the $x$ axis from left to
    right. The symbols $x_n$ enumerate horizons (zeros of $B(x)$) from left
    to right: $x_1 < x_2 < \ldots$. The diagram for B3 occupies the whole
    plane except the dashed triangles. The diagram for B2 cannot be placed
    in a plane since there are overlapping regions, as shown in the picture:
    the $R_3$ region drawn below has a flat infinity ($x\to \infty$) on
    the right, while in another $R_3$ region, drawn turned up, such an
    infinity is on the left.}

\label{fig-BV1} \end{figure}

  It is of interest that in the narrow range of $q$ in which the behavior of
  $B(x)$ drastically changes, the potential $V(x)$ changes very little.
  We also notice that at large $|x|$ the potential takes small positive
  values. It is not by chance since in the general case (\ref{V_x}) $V(x)$
  behaves at large $x$ as follows:
\beq
        V(x) = \frac{4m}{5x^5} - \frac{2 q^2}{3 x^6} + O(x^{-7}).
\eeq

\subsection {Asymmetric configurations}

  Concerning asymmetric configurations, it is natural to expect a critical
  behavior, i.e., transitions between different types of models, at values
  of $m$ and $q$ close to those appearing in Fig.\,2 (but certainly with
  $p \ne 0$). This idea is confirmed by a direct inspection, and Fig.\,2
  (left) shows the corresponding five modes of the behavior of $B(x)$
  at $m = 0.725$:

\begin{description}
\item [B1:]
    A black universe (M-dS) with a single simple horizon.
\item [B2:]
    A black universe (M-dS) with two horizons (simple and double).
\item [B3:]
    A black universe (M-dS) with three simple horizons.
\item [B4:]
    A regular extremal \bh\ (M-AdS) with a double horizon, asymptotically
    AdS at the far end ($x\to -\infty$).
\item [B5:]
    A \wh\ (M-AdS), asymptotically AdS at the far end ($x\to -\infty$).
\end{description}

  The shape of the potential $V(x)$ (Fig.\,3, right) corresponds to \eq
  (\ref{as-V}): it is certainly zero at the flat asymptotic and is of the
  opposite sign to that of $B$ at the other end.

\begin{figure}
\centering
    {\includegraphics[width=8cm]{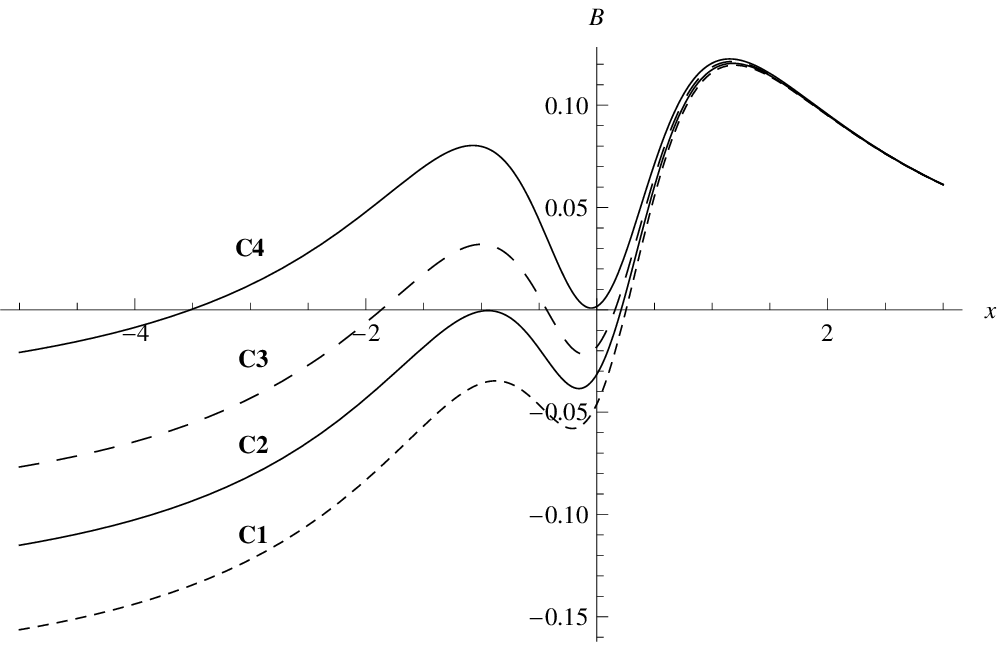}}\ \ \
    {\includegraphics[width=8cm]{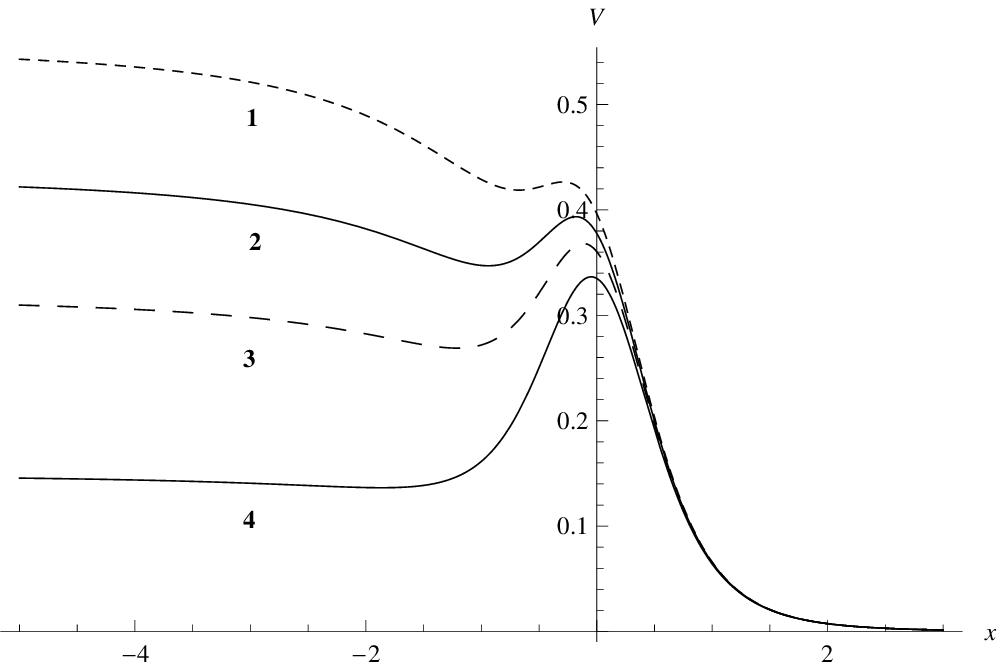}}\\[5pt]
    {\includegraphics[width=6cm]{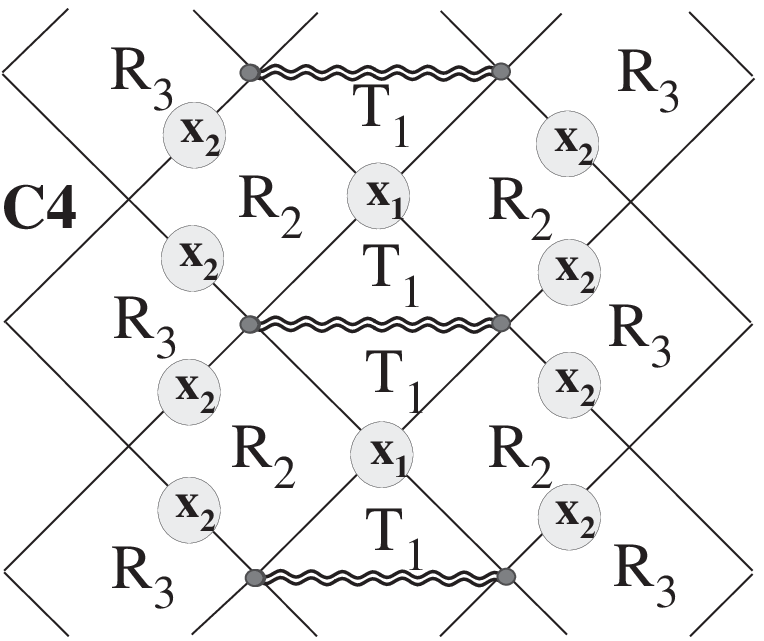}}\
\caption{\small Plots of $B(x)$ (top, left) and $V(x)$ (top, right) for
     asymmetric configurations with the parameter values
     $m= 0.7$ and $q = 0.806,\ 0.8086,\ 0.811,\ 0.8148$ for curves C1--C4,
     respectively. In the Carter-Penrose diagram for C4 shown below, the
     notations are the same as in Fig.\,2.} \label{fig-BV1}
\end{figure}

  A somewhat different picture is observed if we slightly move down the mass
  and charge values, see Fig.\,3 corresponding to $m = 0.7$. A qualitatively
  new feature as compared to Fig.\,2 is that the function $B(x)$
  corresponding to a double horizon between two R-regions (curve C4) has a
  negative limit as $x\to -\infty$.  As a result, it is a \bu\ model instead
  of an M-AdS regular BH. Accordingly, the global causal structure
  characterized by the Carter-Penrose diagram is quite different (Fig.\,3,
  bottom).

\begin{figure}
\centering
    {\includegraphics[width=6.5cm]{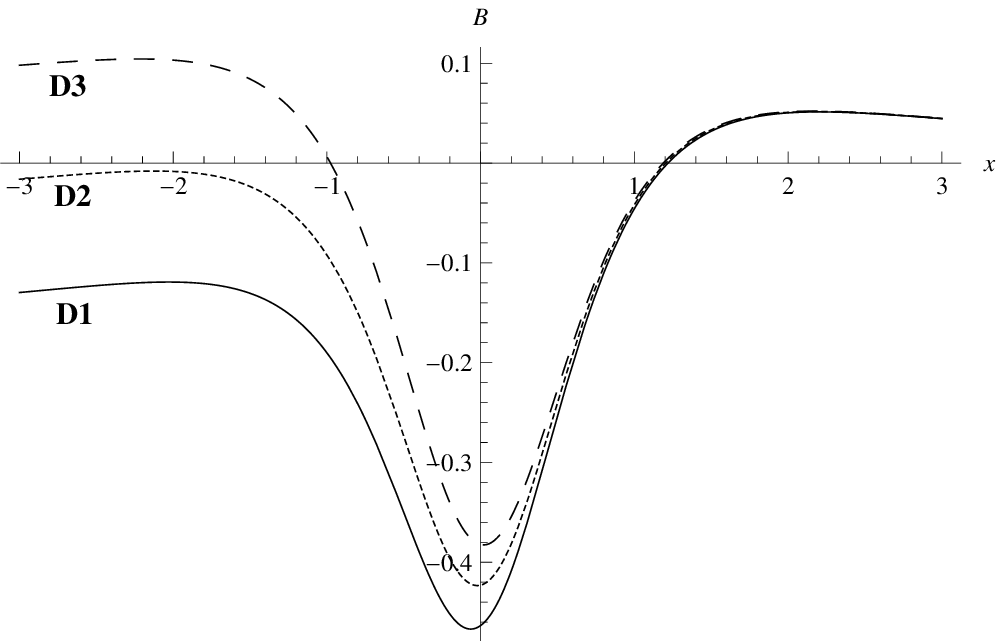}}\
    {\includegraphics[width=6.5cm]{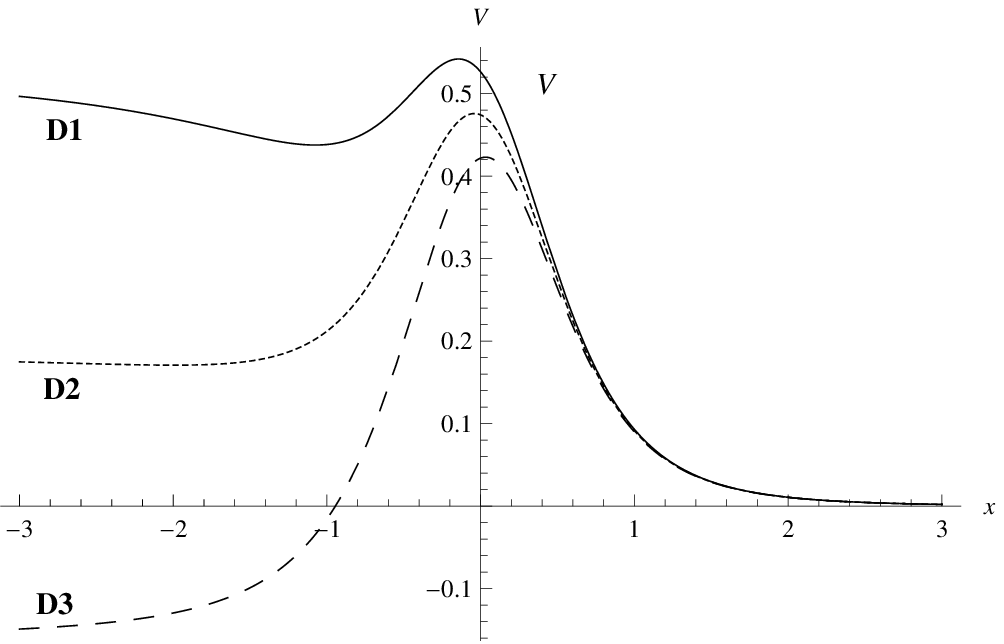}}\
    {\includegraphics[width=2.7cm]{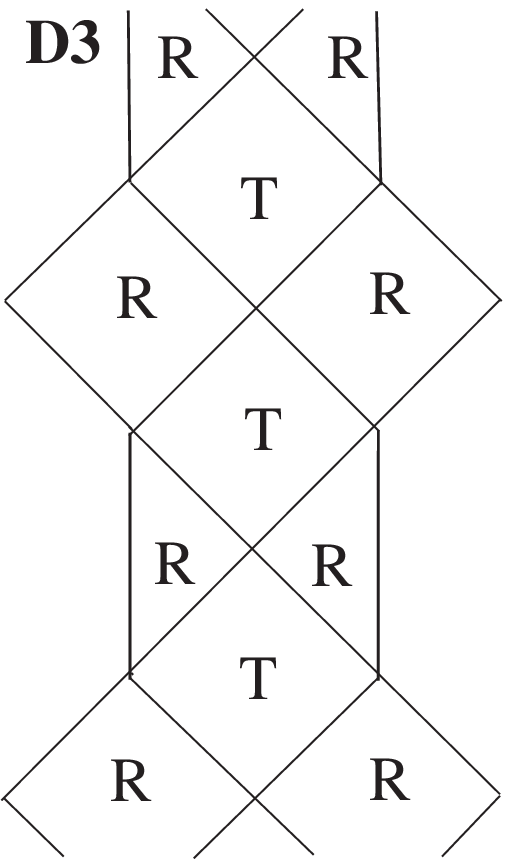}}
\caption{\small Plots of $B(x)$ (left) and $V(x)$ (middle) for asymmetric
        configurations with the parameter values
    $m =1$ and $q=0.968$ (curve D1), $q=0.974$ (curve D2) and $q=0.98$
    (curve D3). The right panel shows the Carter-Penrose diagram for D3.
        }       \label{fig-BV1}
\end{figure}

  Less diverse is the solution behavior at larger values of the
  parameters, as exemplified in Fig.\,5:

\begin{description}
\item [D1, D2:]
    Black universes (M-dS) with a single simple horizon,
\item [D3:]
    Regular \bhs\ (M-AdS) with two simple horizons.
\end{description}

  The curve D3 corresponds to one more type of global causal structure: the
  Carter-Penrose diagram (Fig.\,4, right) is the same as for a non-extremal
  \RN\ BH, but instead of a \RN\ central singularity we have an AdS
  infinity.

\begin{figure}
\centering
    {\includegraphics[width=7.5cm]{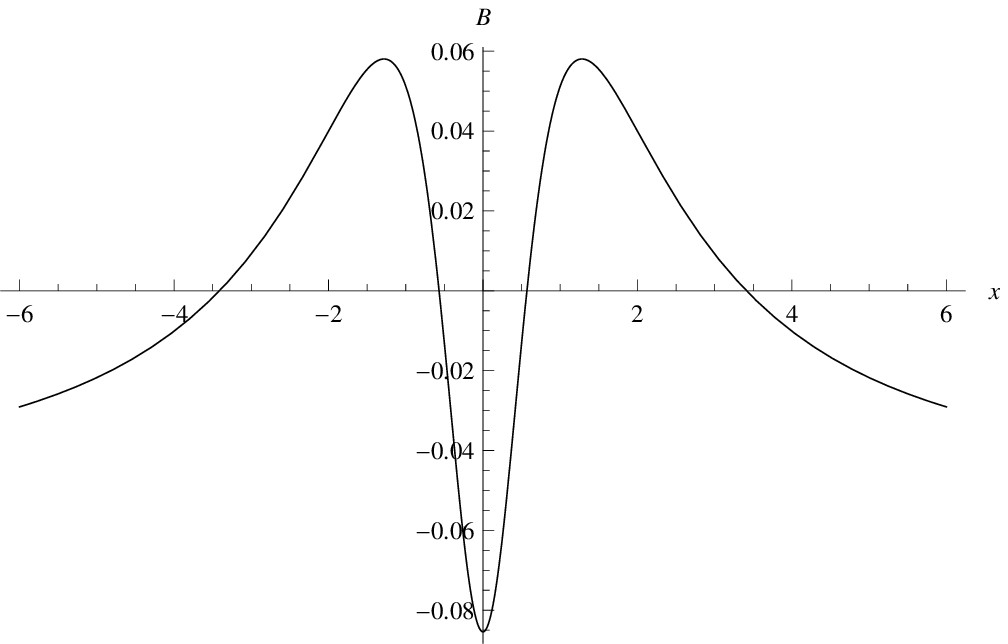}}\qquad
    {\includegraphics[width=7.5cm]{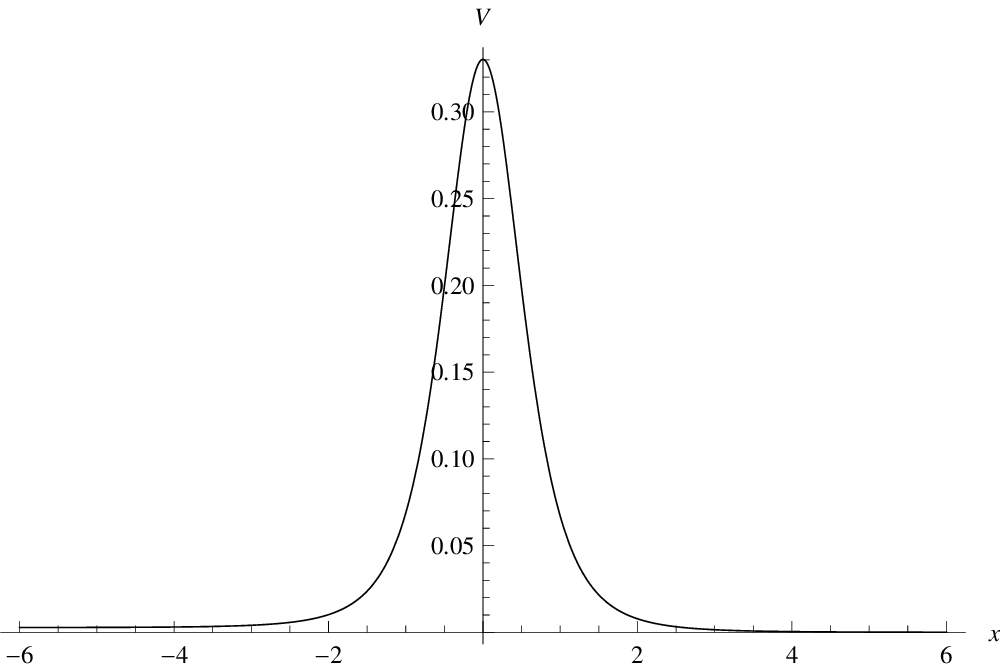}}\\[8pt]
    {\includegraphics[width=7.5cm]{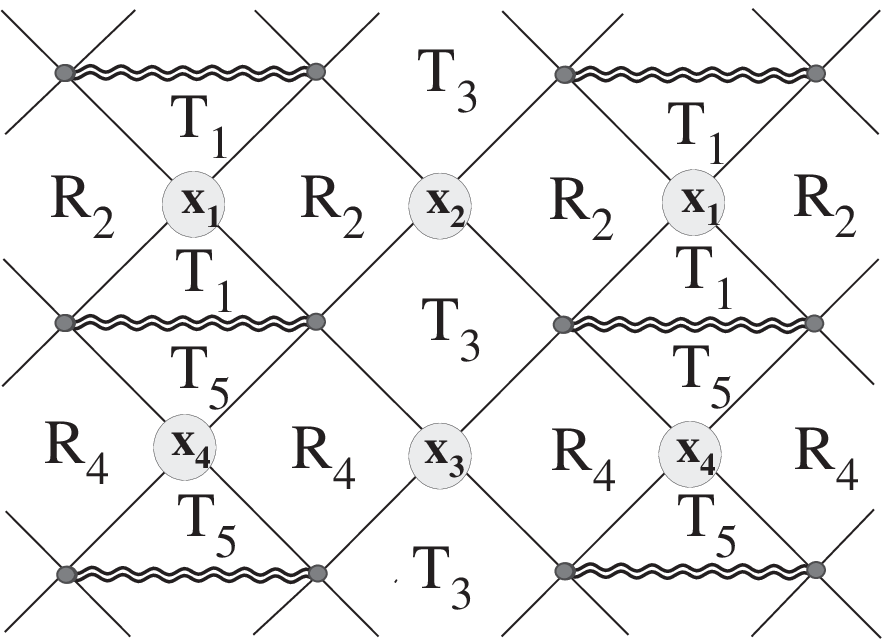}}
\caption{\small A symmetric configuration (dS-dS) with four simple
    horizons, $m = 0.739$, $q= 0.84$, $B_0 = -3\pi m/2 + \pi^2 q^2/4 -0.05$.
    Plots of $B(x)$ (top, left) and $V(x)$ (top, right) and the
    Carter-Penrose diagram (bottom), with the same notations as before. The
    diagram occupies the whole plane except the segments shown by wavy lines
    (de Sitter infinities).  }
\end{figure}

  So far we have been assuming that the space-time is \asflat\ as $x\to
  \infty$.  It is clear that if we abandon this assumption, then the number
  of possible qualitatively different globally regular configurations in
  the scalar-electrovacuum system under consideration will be still larger.
  To see how they can look, let us note that in \eq (\ref{B}) the constant
  $B_0$ is additive. Therefore, changing $B_0$, we simply move up or down
  the plot of $B(x)$, thus changing the asymptotic behavior and the number
  and nature of horizons in our model. For instance, if we slightly move
  down the curve A3 in Fig.\,2, we obtain a configuration with two de Sitter
  asymptotics (dS-dS), separated by four simple horizons, see Fig.\,5.

  In the same way it is easy to obtain a number of other configurations with
  dS and AdS asymptotic behaviors.

\section {Discussion}

  Scalar-vacuum configurations with a self-interacting phantom scalar field
  have been considered in \cite{pha1, pha4} (see also references therein);
  they included M-M and M-AdS \whs\ and black universes. In the present
  paper, we have obtained similar models with an electromagnetic field added
  and found that its inclusion leads to a greater diversity of qualitatively
  different configurations. More specifically, even being restricted to
  solutions which are \asflat\ as $x \to \infty$ and have $m \geq 0$, we
  have found as many as 10 types of models, classified by the types of
  asymptotic behavior and the number and nature of horizons. At zero charge
  $q$ we return to the situation discussed in \cite{pha1, pha4}, with only
  two configuration types: M-M \whs\ with $m=0$ (its analogue is represented
  here by the curve A1) and black universes with a single simple horizon
  (a similar behavior of $B(x)$ is shown here, e.g., by the curves B1 and
  C1); also, M-AdS \whs\ were obtained there but only for $m < 0$. As
  already mentioned, the reason for such a narrow choice is that in a pure
  scalar-vacuum system the field equations forbid the function $B(u)$ to
  have a regular minimum \cite{vac1}.

  Different types of regular configurations obtained here, which are
  \asflat\ as $x\to \infty$ and have a nonnegative Schwarzschild mass, are
  summarized in the table.

\begin{table}[h,t]
\centering
\caption{Types of \asflat\ solutions with $m > 0$}
\medskip
\begin{tabular}{|c|l|l|}
\hline
Solution type & Configuration type, asymptotics & Horizons: number, order $n$\\
    (curve number) & ($x \to+\infty$) ---  ($x\to-\infty$) &
    [disposition of R- and T-regions]\\[5pt]
\hline
   A1,A2 & M - M \wh & none [R]\\
\hline
   A3 & M - M extremal \bh & 1 hor., $n=2$ [RR]\\
\hline
   A4 & M - M \bh & 2 hor., $n=1$ (both) [RTR]\\
\hline
   B1, C1, C4, D1, D2 & M - dS \bu & 1 hor., $n=1$ [TR]\\
\hline
   B2, C2 & M - dS \bu & 2 hor., $n=2$ and $n=1$ [TTR]\\
\hline
   C4 & M - dS \bu & 2 hor., $n=1$ and $n=2$ [TRR]\\
\hline
   B3, C3 & M - dS \bu & 3 hor., $n=1$ (for all), [TRTR]\\
\hline
   D3 & M - AdS \bh & 2 hor., $n=1$ [RTR]\\
\hline
   B4 & M - AdS extremal \bh & 1 hor., $n=2$ [RR]\\
\hline
   B5 & M - AdS \wh & none [R]\\
\hline
\end{tabular}
\end{table}

\begin{figure}
\centering
    \includegraphics[width=12cm]{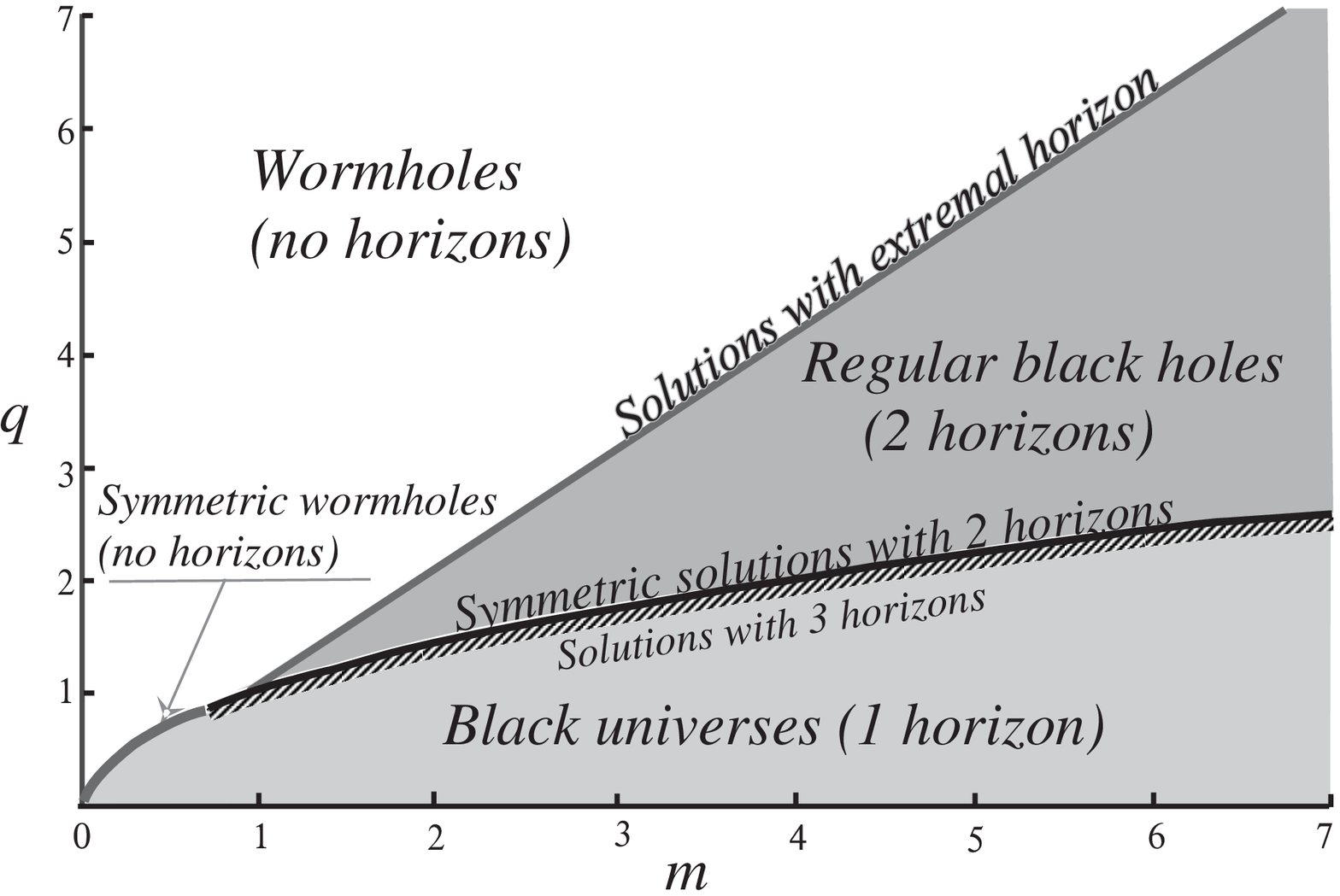}\
    \raisebox{1cm}{\includegraphics[width=5.5cm]{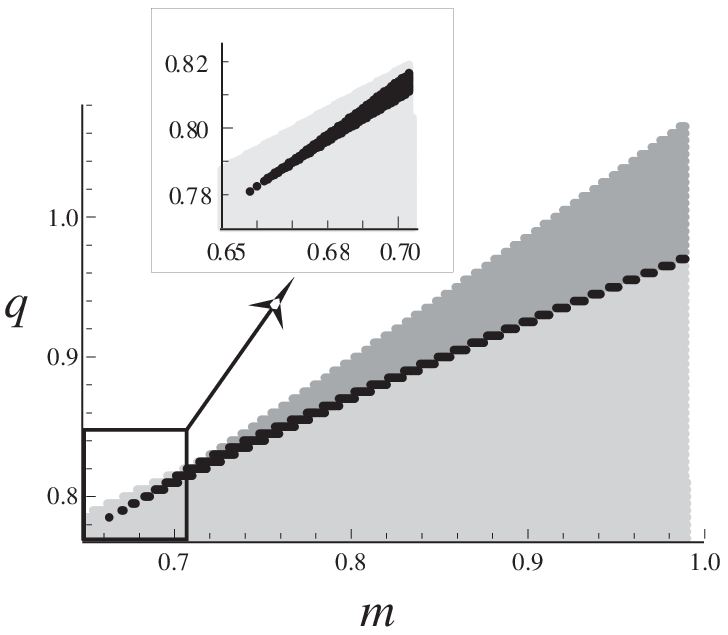}}
\caption{\small Map of \asflat\ solutions with $m > 0$. Right panel:
    a domain of interest enlarged}
\end{figure}

  The general landscape of our solutions with a flat infinity on the right
  end and a positive Schwarzschild mass can be characterized by a map drawn
  in the $(m,q)$ plane (Fig.\,6). It can be seen that the most generic are
  M-AdS \wh\ solutions and black universes with a single simple horizon:
  they exist for all values of $m$ and are actually the same types of
  solutions that have been obtained before with $q=0$ \cite{pha1}. One more
  generic type of models is formed by regular BHs with two horizons, which
  appear only with sufficiently large charges. Solutions with extremal
  horizons appear on separatrices between the main domains on the plane,
  while solutions with three simple horizons are also generic since they
  occupy a certain area on the plane, but this area is actually a very
  narrow band, it is almost invisible if we do not specially adjust the
  scale, see the right panel in Fig.\,6.

  We conclude that the present field system creates quite a number of
  diverse models, making us substantially widen the list of possible regular
  BH configurations as compared, e.g., with \cite{pha4}. Such models can be
  of interest both as descriptions of local objects (\bhs, \whs) and as a
  basis for building singularity-free cosmological scenarios. An important
  feature of such cosmologies, different from the great majority of
  nonsingular models described in the literature, is that the cosmological
  expansion starts from a Killing horizon (this phenomenon can be termed a
  Null Big Bang \cite{br-dym07, kb-zasl07}) beyond which, in the absolute
  past, there is an \asflat\ static region. There is another kind of
  configurations with a Null Big Bang where a static region, instead,
  contains a regular center \cite{br-dym07, br-dym12}; as in the present
  paper, the models described there can possess multiple horizons and have a
  de Sitter asymptotic behavior at late times.

  An important point concerns the value of the global magnetic field that
  exists in our Universe if it can be described by a model more or less like
  ours. Let us use the (probably) most conservative estimate, according to
  which a lower limit on the magnetic field strength is $B \gtrsim 10^{-18}$
  G \cite{dermer11}. On the other hand, the present scale factor $a_0
  \approx 10^{28}$ cm approximately coincides with the quantity $r(u)$ in
  the metric (\ref{ds2}). Therefore, since $B \propto r^{-2}$ [see
  (\ref{F_mn})], we can roughly estimate the field values at earlier stages
  of the evolution. For instance, at recombination ($a/a_0 \sim 10^{-3}$),
  when the electromagnetic radiation decoupled from matter, the magnetic
  field was still weak enough, $B \sim 10^{-12}$ G, but at the stage of
  baryogenesis ($a/a_0 \sim 10^{-12}$) it was of the order of $10^6$ G.

  Observations of the cosmic microwave background (CNB) show that our
  Universe is highly isotropic: at recombination, the degree of anisotropy
  did not exceed $10^{-6}$. If this condition holds, \sph\ models like ours
  (belonging to the \KS\ class), being anisotropic by construction, can
  still conform to observations \cite{craw}.

  The above condition constrains the global magnetic field strength allowed
  by the observed CMB isotropy. The CMB energy density $\rho_{\rm CMB}$ and
  that of the the magnetic field, $\rho_{\rm magn}$, are both proportional
  to $r^{-4}$ as long as the Universe is approximately isotropic, hence
  their ratio is constant and is the same at recombination and at present.
  But at present $\rho_{\rm CMB} \sim 10^{-33}$ \dens, hence we should
  require $\rho_{\rm magn} \lesssim 10^{-39}$ \dens, which in turn means
  $|B| \lesssim 10^{-8}$ G. We see that this condition is easily satisfied
  by the fields under consideration.

  An upper limit on the classical magnetic field description seems to follow
  from the work of Ambjorn and Olesen \cite{amb90} who pointed out that the
  Weinberg-Salam model of electroweak interactions shows an instability at
  $B \gtrsim 10^{24}$ G, connected with emergence of a tachyonic mode.
  If at present $|B| \sim 10^{-18}$ G, then the maximum admissible value of
  $B$ corresponds to $r \approx 10^7$ cm $= 100$ km --- it is the minimum
  admissible value of $b = \min r(u)$ in our models.

  The possible viability of models like those considered in this paper depends
  on their stability under various kinds of perturbations. Most of the known
  scalar-vacuum \wh\ and \bku\ solutions proved to be unstable under radial
  perturbations \cite{gonz, stab1, stab2}, and it is of interest to find out
  whether or not they can be stabilized by electric or magnetic fields. We
  hope to consider this problem in the near future.

\subsection*{Acknowledgments}

    This work was supported in part by NPK MU grant at PFUR and by FTsP
    ``Nauchnye i nauchno-pedagogicheskie kadry innovatsionnoy Rossii'' for
    the years 2009-2013.

\end{document}